# Hyperbolic Metamaterials via Hierarchical Block Copolymer Nanostructures

*Irdi Murataj, Marwan Channab, Eleonora Cara, Candido F. Pirri, Luca Boarino, Angelo Angelini, Federico Ferrarese Lupi\**

I. Murataj, M. Channab

Dipartimento di Scienza Applicata e Tecnologia, Politecnico di Torino, C.so Duca degli Abruzzi 24, Turin 10129, Italy

Advanced Materials and Life Sciences, Istituto Nazionale di Ricerca Metrologica (INRiM), Strada delle Cacce 91, 10135, Turin, Italy

Dr. E. Cara, Dr. Luca Boarino, Dr. Angelo Angelini, Dr. F. Ferrarese Lupi

Advanced Materials and Life Sciences, Istituto Nazionale di Ricerca Metrologica (INRiM), Strada delle Cacce 91, 10135, Turin, Italy

Prof. C. F. Pirri

Dipartimento di Scienza Applicata e Tecnologia, Politecnico di Torino, C.so Duca degli Abruzzi 24, Turin 10129, Italy

E-mail: f.ferrareselupi@inrim.it (F. Ferrarese Lupi, corresponding author)

Hyperbolic metamaterials (HMMs) offer unconventional properties in the field of optics, enabling the opportunity for confinement and propagation of light at the nanoscale. In-plane orientation of the optical axis, in the direction coinciding with the anisotropy of the HMMs, is desirable for a variety of novel applications in nanophotonics and imaging. Here, a method for creating localized HMMs with in-plane optical axis based on block copolymers (BCPs) blend instability is introduced. The dewetting of BCP thin film over topographically defined substrates generates droplets composed of highly ordered lamellar nanostructures in hierarchical configuration. The hierarchical nanostructures represent a valuable platform for the subsequent pattern transfer into a Au/air HMM, exhibiting hyperbolic behavior in a broad wavelength range in the visible spectrum. A computed Purcell factor as high as 32 at 580 nm supports the strong reduction in the fluorescence lifetime of defects in nanodiamonds placed on top of the HMM.

## 1. Introduction

In the last decades, the fast development of nanofabrication techniques has progressively enabled a fine and precise control of the material properties. In the field of optics, the ability to fabricate deep subwavelength structures, the so-called meta-atoms,[1] opened up the possibility to design the individual optical responses of the building blocks of artificial metamaterials instead of addressing the bulk properties.[2] As an example, it is well-known that hyperbolic metamaterials (HMMs), based on their constituent meta-atoms, can exhibit a hyperbolic isofrequency surface of the optical modes in a broad wavelength range, including visible frequencies, due to the strong anisotropy of the electrical permittivity.[3] These materials show some exotic optical properties that are unattainable in bulk structures, such as negative refraction[4] and a broadband enhancement of the local density of optical states (LDOS),[5] with the advantage that the optical behavior can be tuned by properly designing the meta-atoms.[6] These unconventional properties open many opportunities for controlling the confinement and propagation of electromagnetic waves at the nanoscale in time and space,[7,8] with intriguing consequences in different fields and applications.[9]

To date, most of the HMMs presented in the literature are realized by sequential growth or deposition of metal/dielectric multilayers[10] or by electrodeposition in nanoporous templates.[11,12] However, these fabrication processes lead to the generation of HMMs whose anisotropy axis lays in the out-of-plane direction and whose high wave-vector modes are buried inside the material and cannot propagate to the far field. As such, single-photon sources (SPSs) placed in proximity of these materials suffer from a low photon extraction decay rate despite the sensible enhancement of their spontaneous emission. This is determined by the poor coupling efficiency with the guided modes of nanofibers or waveguides, limiting as a consequence the feasibility of more complex HMM-based on-chip photonic devices and circuitries.[13,14] Recently, it has been theoretically demonstrated that the on-chip photon extraction of SPSs can be enhanced by properly tilting the optical axis of the HMM with respect to the normal vector of the end-facet of a waveguide.[15] In order to validate the theoretical predictions, it is therefore required the fabrication of HMMs in which the anisotropy axis lays in the in-plane orientation.

As argued by High et al.,[4] 2D HMMs with in-plane optical axis would provide several advantages in applied fields such as integrated photonic circuits and subdiffraction imaging. However, 2D HMMs, as well as subwavelength structures to couple far field radiation to the high wave-vector modes, rely on complex fabrication processes as focused ion beam (FIB) milling and electron beam lithography (EBL) unsuitable for large-scale scalability. The realization of HMMs working at visible wavelengths in the in-plane configuration is therefore still challenging due to technological constraints related to the pattern definition and material deposition at the sub-20 nm level.

Among the various nanopatterning methods, self-assembly (SA) of block copolymers (BCPs) has emerged as a cost-effective and scalable bottom-up approach, with a high throughput and able to provide highly dense and periodic patterns at the nanometric scale.[16-18] The extensive research over the last years has proved self-assembly of BCPs as a valid alternative to conventional nanopatterning techniques for many different applications including optics,[19,20] biosensing,[21] lithography,[22,23] and electronics.[24,25]

Self-assembled BCPs on unpatterned substrates, however, exhibit highly defective morphologies that limit their technological exploitation. Lots of efforts have been made in order to address the nanofeatures ordering by homopolymer blending[26-28] and implementing the BCP self-registration into pre-patterned substrates *e.g.* directed self-assembly (DSA) or self-registered self-assembly (SRSA).[29-34]

As reported in our previous work,[35] the dewetting process determined by BCP/Homopolymer blend instability generates highly ordered lamellar nanofeatures perpendicularly oriented to the substrate in a single-grain configuration, representing a promising method for fabricating large-scale templates for metamaterials with in-plane optical axis. However, its practical implementation is prevented by the random disposition of the dewetted droplets over unpatterned surfaces. Although a number of hierarchical BCP patterns obtained by sequentially self-assembled BCP,[36] shear-induced alignment,[37] soft graphoepitaxy[38] and thickness-modulated BCP films[39] are reported in the literature, the realization of plasmonic substrates and metasurfaces after metal loading or deposition[40-42] still remains challenging due to the relatively low aspect ratio of the metallic nanofeatures.

Here, we present a novel approach for the fabrication of HMMs having the anisotropy axis in the in-plane direction based on the hierarchical dewetting of BCP blends. The spatial placing and the control of the shape and dimensions of micrometer-sized dewetted features were achieved by means of large-area topographically defined substrates obtained with conventional direct laser writing lithography. By directing the dewetting process over specifically designed motifs, we exerted the control on the nanoscale ordering, obtaining isolated droplets with lamellar nanostructures in a single-grain configuration. Finally, the pattern transfer, assisted by a flexible UV cured resin, revealed a lamellar Au/air HHM replicating the hierarchical BCP template morphology with sub-20 nm features. Effective medium approximation was used to assess the metamaterial response revealing a hyperbolic dispersion spanning over the entire visible wavelength range and a broadband Purcell factor. Experimental evidence of a strong reduction in the fluorescence lifetime of nitrogen-vacancy (NV) centers in nanodiamonds supports the hyperbolic behavior of the metamaterial.

## 2. Results

### 2.1. Dewetting Process on Unpatterned Substrate

Blending BCPs with low molecular weight homopolymers enhances the lateral ordering of self-assembled nanometric features up to one order of magnitude higher compared to unblended BCP.[28,43] At the same time, a reduced effective glass transition temperature imposed by the so-called "wet brush" regime[28] significantly affects the thermal stability of the BCP film. As a result, a low-temperature dewetting process, typically occurring in ultrathin (sub-10 nm) films,[44] takes place on relatively thick BCPs.[35]

The stochastic nature of the dewetting process induced in unstable BCP blends generates droplets with a great variety of contours and dimensions randomly arranged on a flat substrate **(Figure 1a)**.

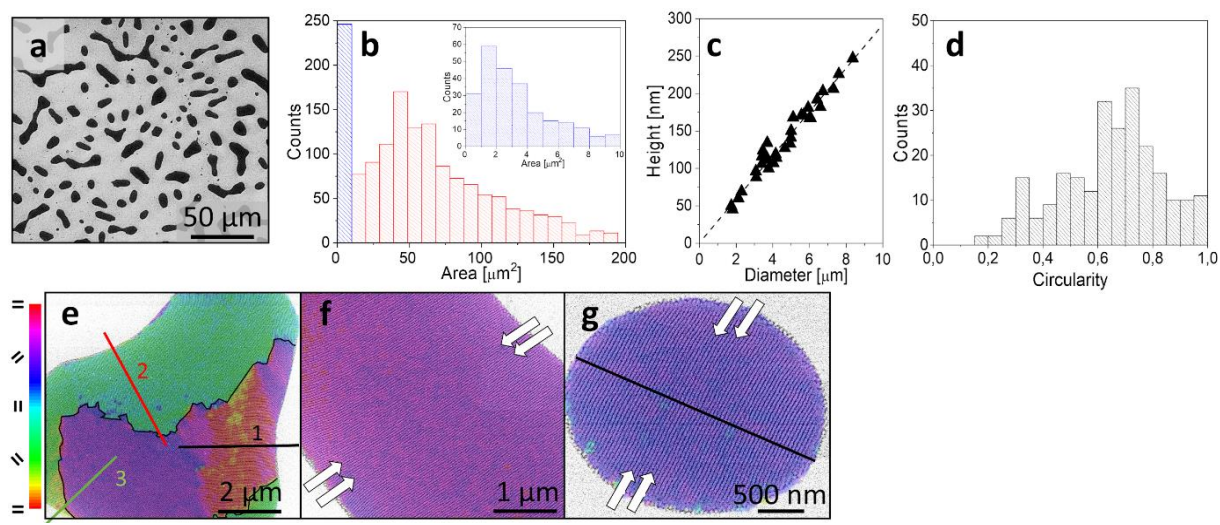

**Figure 1**. a) SEM micrograph representing BCP blend dewetting over an unpatterned substrate and relative droplet area distribution with inset of droplets with sub-10 µm² area (b) (blue columns). c) Droplets heights as a function of the diameter and their circularity (d). e) SEM micrograph of a droplet with irregular shape, overlapped to the false-color map describing the orientation of the lamellae along adjacent edges (indicated by the black, red and green lines).

Similar analysis was performed on elongated droplets with parallel dewetting fronts (indicated by the white arrows) (f) and elliptically-shaped droplets (g). Topographic characterization, performed by scanning electron microscopy (SEM), describes the formation of two main droplet distributions: one broad distribution ranging from 10 to 200 $\mu m^2$ (highlighted in red in Figure 1b) and a second one composed of sub-10 $\mu m^2$ area droplets (blue stripe in Figure 1b). The inset graphic in Figure 1b represents an extended view of the sub-10 $\mu m^2$ distribution. While the presence of two distinct types of droplet populations was previously reported in unstable lamellar-forming poly(styrene-*block*-para- methylstyrene) BCP stored under toluene vapor conditions,[45] its origin is still unclear.

The morphological characterization, performed by optical profilometry, reveals a linear dependence of droplets height on their diameter (Figure 1c). Such behavior implies an intrinsic limit to the use of the dewetting process as lithographic tool, namely the fixed ratio between maximum thickness achievable and the dimension of the patterned area. Moreover, the wide distribution of circularities observed on a flat unpatterned surface (Figure 1d) imposes further restrictions on their systematic technological exploitation.

According to the described picture and considering the tendency of the BCP nanostructures to spontaneously align along the thickness gradients in thickness-modulated films,[46] it is plausible to assume that the extreme variability of the dewetted features could affect the ordering of the nanostructures. To assess the impact of droplets morphology on the nanoscale ordering, we systematically analyzed the SEM micrographs of dewetted features, overlapping the corresponding color maps describing the orientation of the lamellae[47] (a better visualization can be found on Figure S1). In droplets with irregular shape, the alignment of the lamellae along adjacent edges (indicated by the black, red and green lines in Figure 1e) generates defects in the inner zone. On the contrary, in more regular and elongated droplets with parallel borders (highlighted by the white arrows in Figure 1f) the alignment along opposite dewetting fronts promotes the formation of a single defectless orientation, spanning across the entire dewetted structure. Similar behaviour is observed in elliptically-shaped droplets, in which highly ordered lamellae, oriented along the direction orthogonal to their major axis, are observed (Figure 1g). According to the presented results, tailoring the shape of the micrometric droplets is a promising pathway for the achievement of single-grain and defectless lamellar configuration, representing an ideal platform for the attainment of HMMs with in-plane optical axis.

## 2.2. Fabrication of Hyperbolic Metamaterial

Different strategies have been explored so far to exert the control over position, size, and shape of dewetted structures, involving the use of chemically or physically patterned substrates[44,48] or by micro-contact and transfer printing.[39,49] In this work, laser-writer lithography and reactive ion etching (RIE) were used for the realization of periodic guiding templates with micrometric dimensions into a Si wafer. These methods represent a simple and fast patterning avenue, suitable for the definition of the micrometric templates and owning sufficient robustness to withstand the subsequent processes involved in the HMM fabrication. The guiding templates consist of periodic arrays of etched squares with 200 nm depth and gap distances between two consecutive squares ranging from 5 μm to 50 μm, resulting in topographical patterns with grid meshes layouts of different dimensions. SEM analysis performed on templates having gap distances above 10 μm revealed the formation of multiple droplets on top of the grid meshes, whose dimensions and shapes roughly replicate those observed on the flat surface (Figure S2). This indicates that wide templates are inadequate to properly guide the dewetted droplets into defined positions and with desired morphologies and dimensions. On the other hand, on templates with gap distances of 10 μm, single droplets can be found located on top of each grid mesh. However, a high average droplet area (above 30 μm$^2$) prevents the self-assembly process to form lamellar nanostructures in a single-grain and defectless configuration. The dewetting process over smaller guiding templates (5 μm gaps) promotes as well the formation of dewetted droplets on top of the grid mesh layout, organized in an hierarchical configuration (Figure 2a). These guiding patterns, however, emerge as the most attractive since they induce the selection of droplets with single-grain lamellar configuration (Figure 2b,c). The defectless lamellar ordering is to be found on the characteristic dimensions of the two distinct sets of droplets with size distributions peaked at 0.5 and 10 $\mu m^2$ (Figure 2d). Moreover, compared to the very broad distribution observed on the unpatterned surface, the presence of topographic constraints determines the formation of droplets with circularities values above 0.8 (Figure 2e).

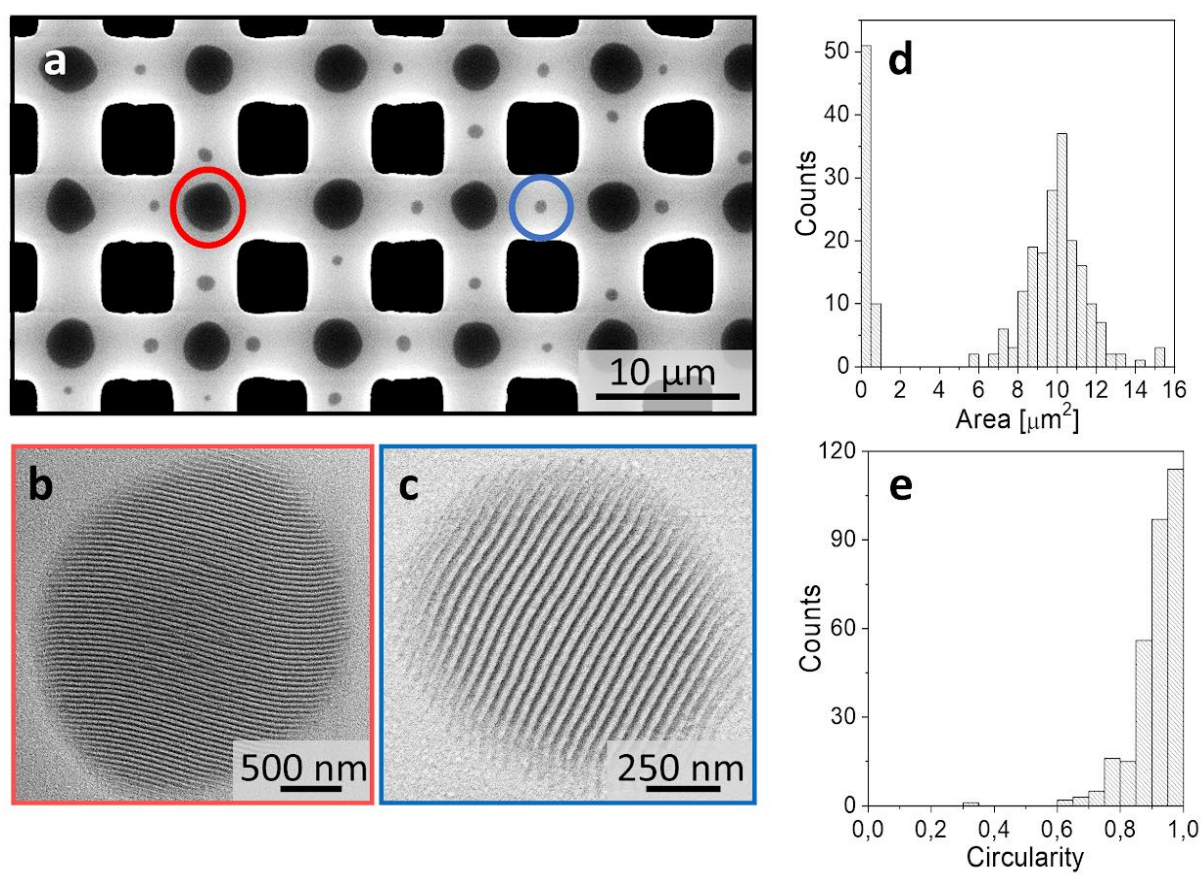

**Figure 2**. a) SEM micrograph of BCP blend dewetting over a large-area topographically defined substrate. SEM images of lamellar nanostructured droplets in a single grain configuration relative to area distribution centered at 10 $\mu m^2$ (b) and at 0.5 $\mu m^2$ (c). d) Area distribution of dewetted droplets over the topographic layout and e relative circularities.

This indicates that the proper tuning of the topographical patterns allows for a fine control over the geometrical characteristics of the dewetted droplets, in terms of dimensions, shape and consequently with desired nanoscale ordering.

The hierarchically organized BCP nanostructures were then used as templates for the fabrication of a Au/air HMM. The entire fabrication process utilized for the realization of the HMM is schematically described in Figure 3a. The dewetted droplets were covered with a 70 nm thick Au layer, followed by a pattern transfer via peeling-off the metallic layer assisted by a flexible UV cured resin, that guarantees a good adhesion on metal and transparency in the visible spectrum. After the pattern transfer process, a perfect replica of the hierarchically organized BCP nanostructures was obtained (Figure 3b). Due to the uniform deposition of Au over the entire micrometric guiding pattern and inside the etched squares, the peeling-off process arises micrometric Au domes spaced apart 5 μm. Each Au dome is interspaced by a lamellar Au/air HMM in a single-grain configuration (Figure 3c) which maintains the characteristic BCP pattern period ($L_0 = 37$ nm) and average lamella width ($d = 18$ nm) (Figure 3d). The so-obtained HMM, presents an optical axis which lays in the in-plane direction normal to the Au/air interface. Such a particular feature, along with the integration into a transparent and flexible substrate, represents an attractive solution for the realization of soft devices, meeting the emerging requirements in flexible and stretchable integrated photonics.[50]

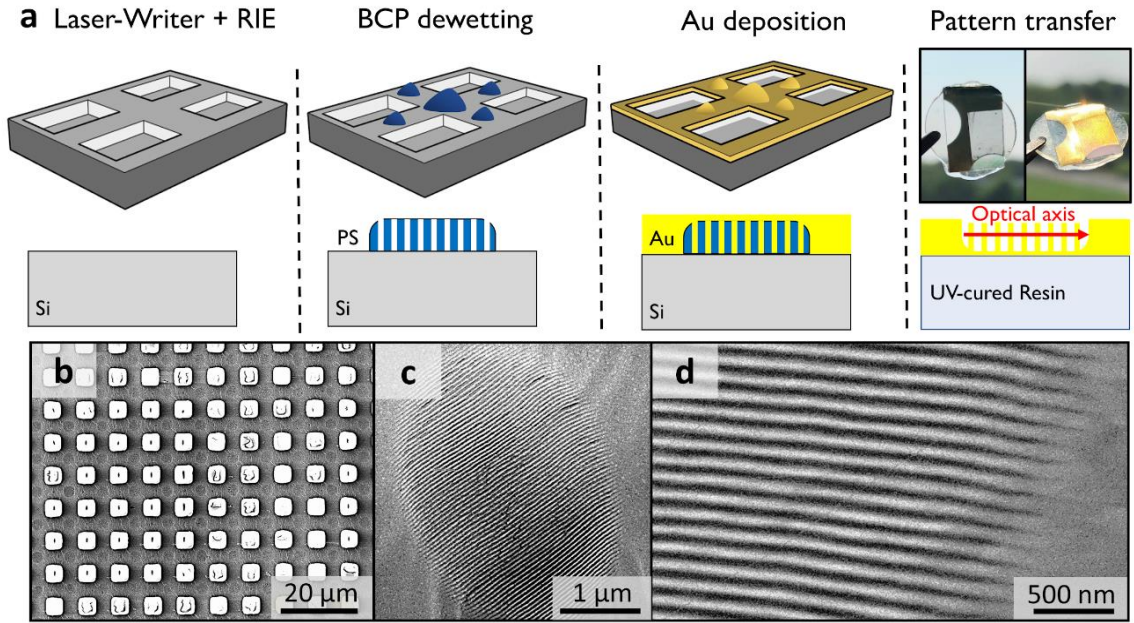

**Figure 3**. a) Schematic fabrication process of the HMM based on hierarchical BCP blend film dewetting over topographical templates and relative SEM micrographs at different magnifications after the pattern transfer process (b,c,d).

## 2.3. Spontaneous Emission Near the Hyperbolic Metamaterial

Numerical analyses, based on the geometrical parameters obtained by SEM characterization shown in the previous paragraph, were performed in order to assess the optical behavior of the fabricated HMM. Figure 4a shows the real and imaginary components of the dielectric permittivity of the HMM in the out-of-plane (orthogonal to the optical axis, blue lines) and in-plane (parallel to the optical axis, green lines) directions, calculated by an effective medium approximation[5] (sketch of the simulation model on Figure S3). The opposite signs of the real components of the permittivity, reveal a hyperbolic dispersion of the HMM over the entire visible range. The reason for such exotic optical behavior is to be found in the anisotropy related to the subwavelength Au lamellar structures, acting as meta-atoms, which originates different responses for the longitudinal and transverse magnetic polarized light.[3,10,51] Moreover, the permittivity components can be further tuned by changing the parameters of the meta-atoms such as the fill factor of the Au with respect to the air (Figure S4).

The hyperbolic dispersion of the optical modes, predicted by the effective medium approximation, leads to high density of electromagnetic states.[1] This generally determines an increase of the Purcell factor associated with the enhancement of the spontaneous emission-rate of single-photon sources coupled to the HMM modes.[52] To quantify such enhancement, the Purcell factor was computed by a commercial software using a finite element method model (Figure 4b). The Purcell factor was estimated as $P/P_0$ where $P$ is the emitted power of a vertically oriented dipole placed 5 nm above the homogeneous effective medium and $P_0$ the power emitted when the dipole is in vacuum. In both cases, the power was calculated as the outgoing flux of the Poynting vector along a circumference with radius of 1 μm in an axisymmetric configuration. The model predicts a broad bandwidth Purcell factor for wavelengths larger than 580 nm, with a 32-fold enhancement at 580 nm.

To experimentally assess the hyperbolic behavior of the HMM, a characterization setup (sketched and described in Figure S5) was employed to measure the luminescence lifetime dynamics of NV centers in nanodiamonds (Figure 4c), previously dispersed over the substrate. Such single-photon emitters were selected since the relative emission spectrum matches the spectral range where the highest Purcell factor was estimated. Time-resolved decay curves, reported in Figure 4d, highlight a strong reduction in the fluorescence lifetime of NV centers in nanodiamonds placed on top of the HMM (blue dots) by about one order of magnitude compared to that observed placing similar nanodiamonds on flat Au (red dots) or bare glass (black dots). The evidence of a strong lifetime reduction is to be considered as the footprint of a larger density of radiative decay channels associated to the hyperbolic dispersion relation.[53] These experimental results, supported by modeling simulations, clearly demonstrate the hyperbolic behavior of the BCP-based HMM, with inherent in-plane optical axis.

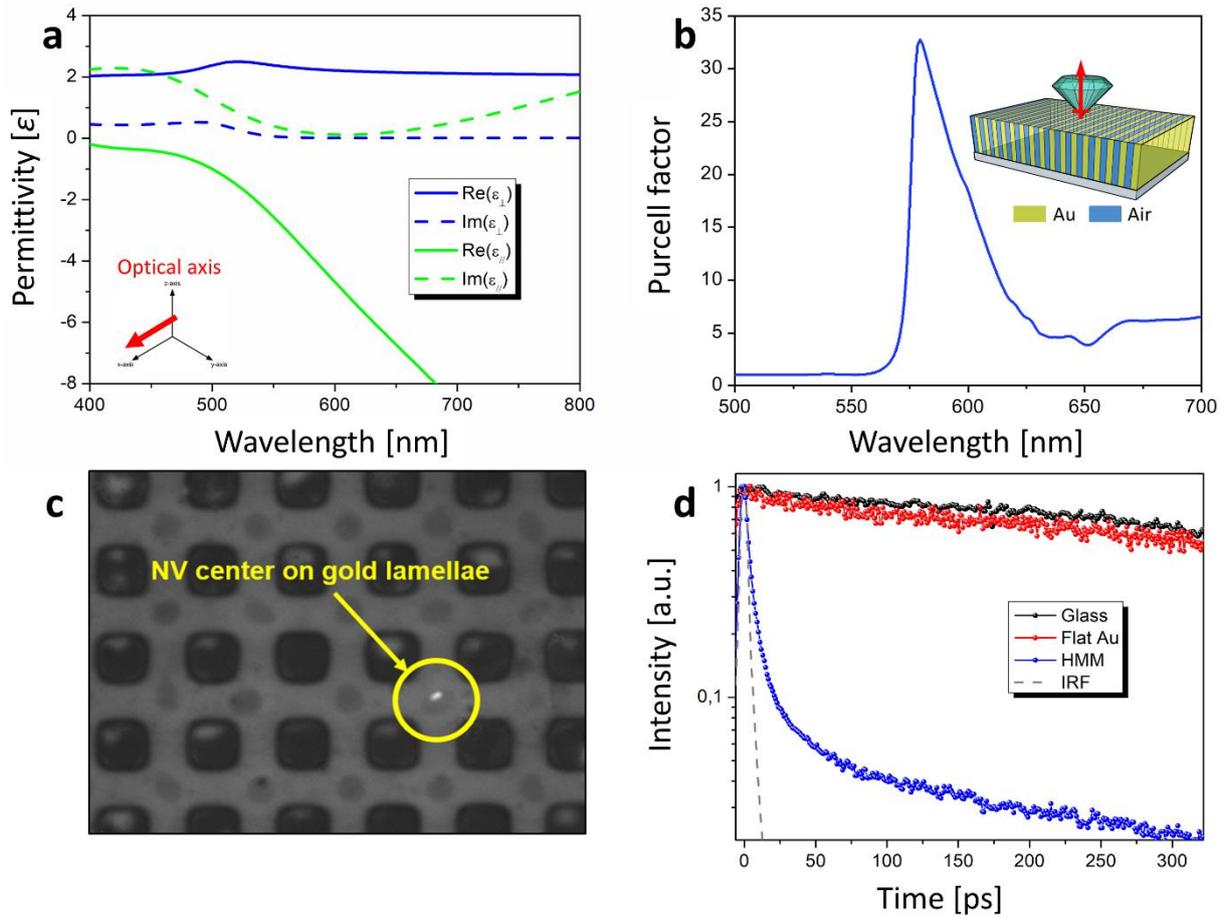

**Figure 4**. a) Real and imaginary components of the effective anisotropic permittivity of the BCP-based HMM calculated in the directions perpendicular (blue lines) and parallel (green lines) with respect to the optical axis. b) Simulated Purcell factor for a vertically oriented dipole (red arrow in the inset) located 5 nm above the HMM. c) White light image of the substrate superposed with the fluorescence image of the excited NV center placed on top of the HMM. d) Lifetime fluorescence measurements for NV centers in nanodiamonds on glass (black dots), flat Au film (red dots) and on top of the HMM (blue dots). The grey dashed line indicates the instrument response function (IRF).

## 3. Conclusion

We showed the fabrication of HMMs with an in-plane optical axis by exploiting the BCP/Homopolymer blend instability. The control over the alignment and ordering of single-grain lamellar BCP nanostructures in hierarchical configuration was achieved by inducing thermally activated dewetting process of BCP blend into topographically defined substrates. The pattern transfer process onto a transparent and flexible substrate, generates a HMM with hyperbolic dispersion in broad wavelength range in the visible spectrum, crucial for the realization of novel smart nanostructured materials for future stretchable and wearable devices.[22] Although a deeper understanding of the interaction between spontaneous emitters and the HMM would require a more extensive characterization, the unconventional optical properties have been demonstrated by a combination of numerical modeling and experimental measurements of the dynamic of the spontaneous emission process. A notable advantage of the proposed fabrication technique relies on the fast and low-cost method to obtain hybrid metal/dielectric substrates patterned at the nanoscale with in-plane optical axis whose orientation can be further tuned through the control over the dewetting process, suitable for on-chip integration. The geometric characteristics of the self-assembled BCP can also be adjusted to tailor the optical environment by finely tuning the fill factor of the metallic nanodomains, the thickness and morphology of the nanopattern and to modify the spectral range through the exploitation of different metals and dielectrics. Moreover, the hierarchical patterning paves the way for an easy fabrication of more complex photonic devices such as photonic hypercrystals[54] and hyperbolic waveguides[52] and provides an additional degree of freedom for the fabrication of graded index elements.[55]

## 4. Methods

*Substrate Neutralization*: Substrate neutralization was performed by grafting process of α-hydroxy ω-Br polystyrene-*random*-polymethyl methacrylate (PS-*r*-PMMA) random copolymer (RCP) (14.60 kg mol$^{-1}$), with styrene fraction (*f*) of 0.59 and a polydispersity index Đ = 1.30,[35] over silicon wafers covered by 1.5 nm thick native oxide layer. A solution of PS-*r*-PMMA (18 mg) in toluene (2 ml) was spin-coated onto the Si wafers at 3000 rpm for 60 s[56] after cleaning and substrate activation by $O_2$ plasma treatment at 130 W for 10 minutes. The grafting process was induced by annealing treatment in a rapid thermal processing (RTP) machine at high temperature ($T_a$ = 290 °C) for an annealing time ($t_a$) of 300 s, followed by sonication in a toluene bath for 6 minutes, obtaining a 7 nm thick grafted RCP layer, measured by spectroscopic ellipsometry (alpha-SE ellipsometer, J.A. Wollam Co.).

*Block Copolymer Blend Deposition*: A polystyrene-*block*-polymethyl methacrylate (PS-*b*-PMMA) (66 kg mol$^{-1}$) with styrene fraction (*f*) of 0.50, polydispersity index Đ = 1.09 was purchased from Polymer Source Inc. and used without further purification. Low molecular weight homopolymer PMMA (3.9 kg mol$^{-1}$) with Đ = 1.18 and PS (3.1 kg mol$^{-1}$) with Đ = 1.09) were synthesized as described in a previous work.[35] A solution prepared by equal amounts of PS and PMMA homopolymers (1.12 mg) blended with BCP (4.5 mg) and dissolved in toluene (2 ml) was spin-coated over RCP-neutralized substrate obtaining a 35 nm thick layer. Subsequent annealing treatment by RTP at 270 °C for 600 s in $N_2$ environment induced the dewetting in the BCP blend film. Finally, a selective removal of polymethyl methacrylate was achieved by exposure of the samples to ultraviolet radiation (5 mW cm$^{-2}$, λ = 253.7 nm) for 180 s followed by isotropic $O_2$ plasma etching (40 W for 30 s).

*Graphoepitaxy Patterns*: Micrometric arrays of different patterns with sizes ranging from 5 to 50 μm and periodicity over a 5 mm x 5 mm area were obtained by direct laser writer lithography (Heidelberg Laser Writer μPG101) using a commercial optical resist (AZ 5214E MicroChemicals GmbH) spin-coated (1 μm thick layer) on Si wafers and subsequently exposed to a laser beam with 15 mW power and wavelength of 375 nm. The development was performed by sample immersion for 40 s in a 1:1 solution of AZ Developer (Merck Performance Materials GmbH) and deionized $H_2O$, and subsequent rinsing in $H_2O$ for 60 s. Reactive ion etching process (RIE) with fluorine-based chemistry ($C_4F_8$ and $SF_6$) was performed in order to transfer the pattern into the Si wafer with a final depth of 200 nm.

*Patterns Transfer:* A 70 nm thick layer of Au was deposited on the samples by a homebuilt RF sputtering system by generating an argon plasma at 100 W at $5 \cdot 10^{-3}$ mbar, yielding a deposition rate of 0.2 nm/s. The pattern transfer was performed by depositing a UV-curing resin (Optical Norland Adhesive 81) on top of the Au layer followed by the mechanical stripping from the polymeric template.

*Morphology Characterization:* BCP templates and Au-stripped films were monitored by FEI Inspect-F field emission gun scanning electron microscope (FEG-SEM) using an Everhart-Thornley secondary electron detector (ETD) and backscattered electron detectors (BSED). The SEM images were analyzed by using ImageJ2 software for the morphological characterization of BCP dewetted features in terms of dimensions and circularity. To this end, the SEM micrographs were binarized with automatic thresholding and the droplet outlines were fitted with ellipses. The ratio between the minor and major axis of the resulting ellipses was used for the calculation of the circularity. The SEM micrographs were processed to evaluate the long-range ordering and orientation of the lamellar nanostructures by using a standard software routine, further details can be found in reference 35. 3D surface profiler (S neox SENSOFAR) and a 150x confocal objective were used for the determination of the droplets heights.

**Supporting Information**

Supporting Information is available from the Wiley Online Library or from the author.

**Acknowledgements**

This project has received funding from the EU-H2020 research and innovation programme under grant agreement No. 654360 having benefited from the access provided by IMB-CNM CSIC (Spain) within the framework of the NFFA-Europe Transnational Access Activity proposal ID840. In particular, we thank Marta Duch Llobera, Xavier Borrisé Nogué and Miguel Zabala Garcia of IMB-CNM CSIC for the kind support. SEM characterization has been performed at NanofacilityPiemonte in INRiM, a laboratory supported by Compagnia di San Paolo Foundation.